# DISSECTING SUBSECOND CADHERIN BOUND STATES REVEALS AN EFFICIENT WAY FOR CELLS TO ACHIEVE ULTRAFAST PROBING OF THEIR ENVIRONMENT

Anne Pierres[1,2,3], Anil Prakasam[4], Dominique Touchard[1,2,3], Anne-Marie Benoliel[1,2,3], Pierre Bongrand[1,2,3,5] & Deborah Leckband[4]

[1]INSERM U600, Marseille, F-13009, France, [2]CNRS, UMR 6212, Marseille F-13009, France, [3] Aix-Marseille Université, Marseille F-13009, France  [4]Department of Chemical and Biomedical Engineering and Department of chemistry, University of Illinois, Urbana-Champaign, Illinois 61801.

[5] *Corresponding author* : Pierre Bongrand, Laboratoire d'Immunologie, Hôpital de Sainte-Marguerite, BP 29, 13274 Marseille Cedex 09 FRANCE. Phone (33) 491 26 03 31, Fax (33) 491 75 73 28, e-mail bongrand@marseille.inserm.fr

## ABSTRACT

Cells continuously probe their environment with membrane receptors, achieving subsecond adaptation of their behaviour [1-3]. Recently, several receptors, including cadherins, were found to bind ligands with a lifetime of order of one second. Here we show at the single molecule level that homotypic C-cadherin association involves transient intermediates lasting less than a few tens of milliseconds. Further, these intermediates transitionned towards more stable states with a kinetic rate displaying exponential decrease with piconewton forces. These features enable cells to detect ligands or measure surrounding mechanical behaviour within a fraction of a second, much more rapidly that was previously thought.

# INTRODUCTION

Recently, atomic force microscopy [4,5], laminar flow chambers [6] or the biomembrane force probe [7,8] were used to study at the single bond level the homophilic association of cadherin molecules. Experiments revealed binding states of about 1 s lifetime between VE-cadherins [4], E-cadherins [5,6], N-cadherins [5] or C-cadherins [8]. The estimated association rate was lower than 100,000 $M^{-1}s^{-1}$ [4]. Here, we took advantage of the laminar flow chamber capacity of accessing much faster events to dissect the interaction between C-cadherin moieties. Microspheres coated with outer domains (EC1-5) of C-cadherins were driven along surfaces coated with the same fragments with very low hydrodynamic forces : the wall shear rate ranged between 4 $s^{-1}$ and 16 $s^{-1}$, subjecting arrested spheres to a viscous drag comprised between 0.25 and 1 piconewton. A single bond was thus expected to maintain a particle at rest during a detectable amount of time. Using a tracking system allowing 20 ms and 80 nm resolution for time and position determination, we analysed sphere motion. This revealed transient intermediates with a lifetime lower than a few tens of milliseconds sensitive to piconewton forces. This may enable cells to probe their environment with unsuspected rapidity.

# MATERIALS AND METHODS

**Cadherin production and purification**. The engineering and production of C-cadherin (CC) fragments with Fc domains fused at the C-termini were described previously [9]. Soluble proteins were purified from cultures of stably transfected chinese hamster Ovary (CHO) cells according to published procedures [9,10].

**Particle and surfaces.** We followed previously published protocols [11]. Particles were streptavidin-derivatized spheres of 2.8 µm diameter coated with biotinylated anti-human Fc antibodies, then an excess of Fc-CC1-5 fusion protein. The chamber floor was made of freshly cleaved mica treated with $NiCl_2$, then cadherin CC1-5 tagged with a terminal hexa-histidin.

**Image acquisition and data processing.** Pixel size was 0.31 x 0.23 $\mu m^2$. Microsphere position was determined with a custom-made software locating the image centroid with 80 nm and 20 ms resolution. The shear rate G (in $s^{-1}$) was derived with high accuracy from the velocity $U_0$ (in µm/s) corresponding to zero average acceleration following [12] :

$$G (s^{-1}) = 1.06 \, U_0 \tag{4}$$

A particle was defined as arrested when it moved by less than $\xi=0.31$ μm during the following time interval $\tau$. A resting particle was defined as departing when it moved by at least $\xi+\Delta$ (=0.62 μm) during the following time interval $\tau$. The true arrest duration $d_t$ was derived from the apparent arrest duration $d_a$ according to [11] :

$$d_t = d_a + \tau - (2\xi + \Delta)/v \tag{5}$$

where v is the average particle velocity. Unbinding plots were obtained by plotting the logarithm of the number of particles remaining bound at time t after arrest versus t.

**Models and data fitting**. Experimental unbinding plots were fitted to theoretical curves according to the following two-step model of bond formation :

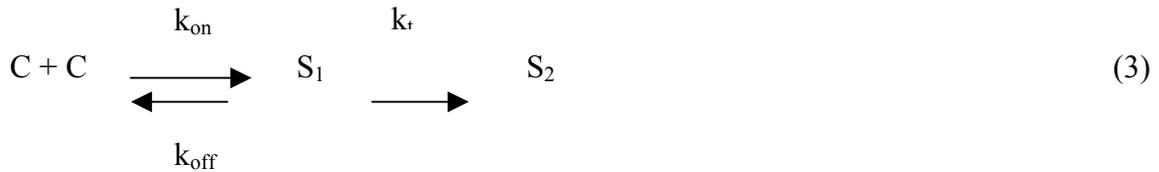

$$C + C \underset{k_{off}}{\overset{k_{on}}{\rightleftarrows}} S_1 \overset{k_t}{\rightarrow} S_2 \tag{3}$$

Assuming that two cadherin moieties will form a transient complex $S_1$ that may undergo either rupture or strengthening to state $S_2$ that will not be dissociated during the observation period of at least 10 seconds. Adjustable parameters used to fit an unbinding plot are the number of binding events $N_0$, the initial dissociation rate $k_{off}$ and the transition rate $k_t$, yielding the following equation :

$$N(t) = [N_0/(k_{off} + k_t)] \, [k_{off} \exp(- (k_{off} + k_t)t) + k_t] \tag{6}$$

Unknown parameters were derived from the number N(t) of particles remaining bound at time 0.2 s, 0.5 s and 1 s. The statistical uncertainty associated with parameter fitting was calculated by estimating at $(N(0.2s) - N(0.5s))^{1/2}$, $(N(0.5s) - N(1s))^{1/2}$ and $N(1s)^{1/2}$ the errors on the number of events falling in intervals [0.2s, 0.5s], [0.5s, 1s] and [1s, ∞] respectively [11].

The information on energy landscapes was derived from the force dependence of kinetic constants using Bells' law [13] :

$$k(F) = k(0) \exp(Fx/k_B T) \tag{7}$$

where k(F) is a kinetic constant for passing a barrier in presence of a force F, x is the distance between the energy minimum and adjacent barrier, $k_B$ is Boltzmann's constant and T is the absolute temperature. Also, we estimated at 0.47 G the force (in piconewton) on a single bond maintaining a particle at rest in presence of a wall shear rate G. As indicated previously, this estimate is only weakly dependent on the bond length.

## RESULTS

Spheres displayed periods of motion with fluctuating velocity interspersed with arrests of variable duration (Fig. 1). As previously shown [12], velocity fluctuations could be quantitatively accounted for by horizontal and vertical brownian motion of microspheres.

We plotted the velocity distribution of microspheres as averaged during 160 ms intervals. Arrests were nearly abolished when experiments were performed in present of calcium chelators (Fig. 2a), thus supporting the interpretation that they were mediated by cadherin-cadherin interactions. Also, in calcium-containing medium, we observed a substantial number of time intervals where particle velocity was lower than expected in absence of bond, but no detectable arrest was found. To check the hypothesis that this finding reflected the occurrence of subsecond arrest periods, we built sets of simulated particle trajectories with i) no bond formation, ii) formation of bonds with a dissociation rate $k_{off}$ of $1s^{-1}$, corresponding to previously reported values, and iii) formation of much more transient bonds with a dissociation rate of $50\ s^{-1}$. As shown in Fig. 2b, experimental histograms might reveal very transient binding states.

Next, we analyzed the quantitative properties of cadherin-cadherin bond formation by studying binding frequency, bond lifetime distribution and force dependence of these parameters. A sphere may be defined as arrested when it moves by less than an arbitrary threshold distance $\xi$ during a time-step of duration $\tau$. A frequently overlooked point [11] is that the true duration $d_t$ of a binding event is dependent on threshold parameters and differs from the apparent duration $d_a$ since :

$$d_t = d_a + \tau - 2\ \xi/v \qquad (1)$$

where v is the mean velocity of moving particles. Also, the minimal duration $d_m$ of detectable binding events is :

$$d_m = \tau - \xi/v \tag{2}$$

As shown of Fig.3 a & b, when Eq.1 is used to correct measured arrest durations, the distribution of lifetime duration of binding events becomes independent of threshold parameters and a threshold-independent number of binding events may be obtained by extrapolation of the curves to time zero. This point is crucial when the effect of shear rate, and accordingly particle velocity, on binding parameters must be quantified.

Now, the detachment of arrested spheres did not exhibit simple monophasic kinetics (Fig. 3b). Curves were fitted with a minimal two-step model of bond formation by assuming that immediately after arrest particles were retained by a transient molecular complex ($S_1$) that might either rupture with kinetic rate $k_{off}$ or transition with rate $k_t$ to a more stable state $S_2$. That strengthening might be due to sequential formation of additional bonds was considered unlikely since i) lateral diffusion was not allowed on particles or surfaces, and all bonds compatible with geometrical constraint should form within the millisecond range, and ii) when the concentration of cadherin molecules used to derivatize surfaces was decreased fourfold, parameters $k_{off}$ and $k_t$ were not significantly altered but binding frequency was decreased as expected if arrests were due to single molecular bonds (not shown).

The key finding of our experiments was obtained by varying the wall shear rate. As shown on Fig. 3c, increasing particle velocity drastically decreased binding frequency. This decrease could not be due to a decrease of bond lifetime, resulting in decreased efficiency of arrest detection since i) this was ruled out by aforementioned correction procedure, ii) bond lifetime was much higher than the detection limit given by Eq.2, and iii) as shown on Fig. 4c, $k_{off}$ was not significantly altered when hydrodynamic forces were increased. In contrast, the bond strenthening rate was markedly decreased when the shear rate was increased.

**DISCUSSION**

The most straightforward interpretation of our results is based on two well-accepted concepts. First, the formation/dissociation of a bimolecular complex often follows a main reaction pathway representing a valley in a mutidimensional energy landscape [14,15]. Second, this pathway may involve a sequence of barriers and basins [15-17]. Further, as first suggested by Bell [13] and checked experimentally [18], the effect of an external force applied on a bimolecular complex is to change the height of these barriers depending on their position, resulting in exponential variation of transition frequencies. Thus, our experimental results may be translated into a topographical description of the external part of the energy landscape of a cadherin-cadherin complex (Fig. 4). To our knowledge this is the first report that a force can slow transition towards an inner binding state, as well as enhance bond rupture [4,15, 18-21]. Several important consequences may be drawn.

First, cells may probe their environment much more rapidly than we might have thought. Indeed, although quantitative information on the association rate of cadherin molecules is rather scarce, experiments done with atomic force microscopy [4], laminar flow chambers [22] and surface plasmon resonance [23] suggested that the contact period required for a cell protrusion of 0.01 $\mu m^2$ area to sense cadherins on an encountered surface might be 100 s or more if the cadherin surface density was about of 100 molecules/$\mu m^2$. However, aforementioned estimates applied to interactions with a lifetime of the order of one second. Our results raise the possibility that much more transient interactions of a few millisecond lifetime might occur during a contact as short as 1 second.

It must be emphasized that substantial evidence supports the view that such transient interactions might be highly relevant to cell function. Indeed, cadherin molecules can influence microfilament organization through catenin molecules [24,25] and/or G protein activation [26]. Since cells may have to reorganize the actin network with subsecond rapidity [1] and receptor-associated GTPases may induce cell responses with similar kinetics [2], subsecond binding states may be highly relevant to cell control. Indeed, this concept is in line with recent data supporting the functional role of ultraweak protein interactions [27].

Second, That the duration of cadherin interaction may be influenced by forces within the piconewton range makes them ideally suited to probe their environment with utmost sensitivity, since this force can be generated by a single actomyosin motor [28,29] and this is sufficient to significantly alter microfilament polymerization [30].

Third, our results emphasize the interest of our experimental setup. Indeed, the average area available for bond formation between a brownian sphere such as were used in our experiments and a ligand-coated surfaces is

$$<A_C> = \frac{\int_{z=0}^{z=p} 2\pi R(L-z)\exp(-\beta Fz)dz}{\int_{a=0}^{\infty} \exp(-\beta Fz)dz} \quad (3)$$

where F is the sedimentation force (i.e. weight minus Archimedes force), L the interaction range, R the sphere radius, and $\beta=1/k_BT$, where $k_B$ is Boltzmann's constant and T is the absolute temperature. Eq(3) yields $<A_C>=0.014$ $\mu m^2$, matching the tip of microvilli. This provides a simple way of mimicking fast probing of their environment by motile cells filopodia.

**Acknowledgements**. This work was supported by NIH GM51338 (DEL) and by the Association pour la Recherche contre le Cancer (PB).

**FIGURE LEGENDS**

**Figure 1. Microsphere tracking. a**, automatic tracking of an individual particle with separation of interlaced images. b, simultaneous recording of position and area. c, velocity fluctuations in 20 ms intervals.

**Figure 2. Velocity histograms. a**, experimental curves. The frequency distribution of particle displacement during 20 ms intervals was determined in absence (circles) or presence (crosses) of calcium. In presence of calcium, there is a peak on the right, corresponding to bound states, and a marked amount of unexpectedly low displacements (arrow). **b**, simulated curves. Dots :

no arrest. Crosses : 112 binding events with dissociation rate 1 s$^{-1}$ for 80 seconds. Circles : 2,272 binding events with a dissociation rate of 50 s$^{-1}$ for a 80 s period.

**Figure 3. Effect of shear on cadherin-cadherin association**. **a**, **b** : Comparison of unbinding plots obtained in the same series of experiments with (b) and without (a) correcting apparent arrest durations as described. Threshold time for arrest duration was 80 ms (squares), 120 ms (crosses), 160 ms (triangles) or 200 ms (circles). **c**, the effect of wall shear rate on binding frequency is shown. The equation of regression line is : lny = - 0.118 x + 0.234. Vertical bar length is twice standard error. **d**, The effect of wall shear rate on $k_{off}$ (triangles) and $k_t$ is shown.

**Figure 4. Suggested reaction pathway for cadherin-cadherin association**. Shown model is consistent with experimental data. The basic hypothesis is that bond formation will involve a passage through a series of transient binding states with increasing stability. State $S_0$ might not be detecable with dissociation rate k. higher than 10 s$^{-1}$. State $S_1$ is a transient binding state with dissociation rate $k_{off}$ towards $S_0$ through a steep barrier and strengthening transition towards state $S_2$ with a rate constant $k_t$. The presence of an external force will change ABCDE into A'B'C'D'E as predicted [13].

Fig. 1

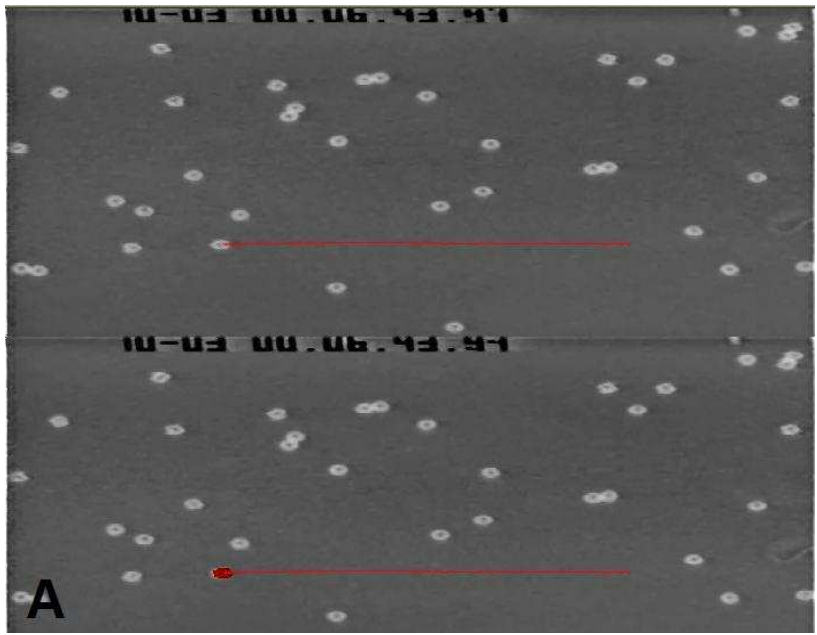

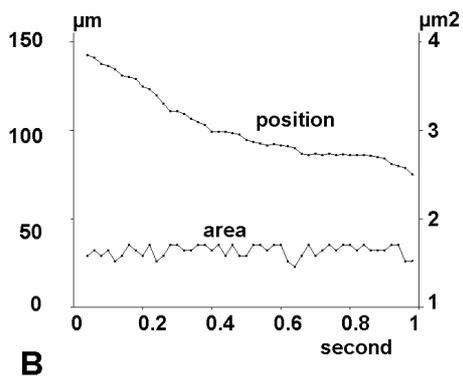

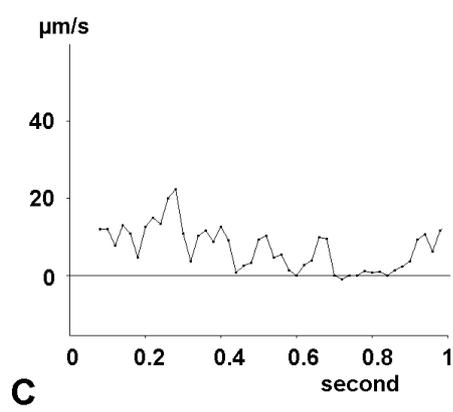

Fig. 2

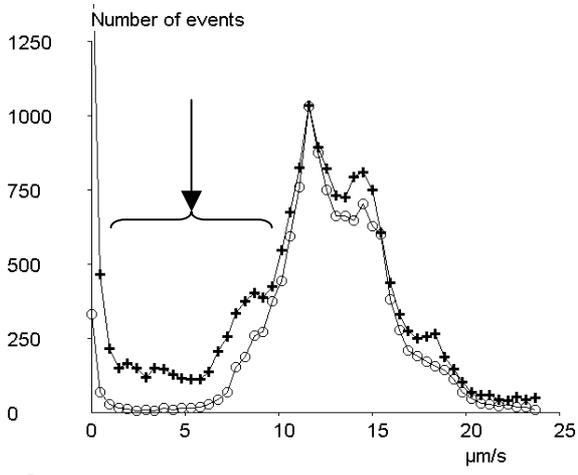 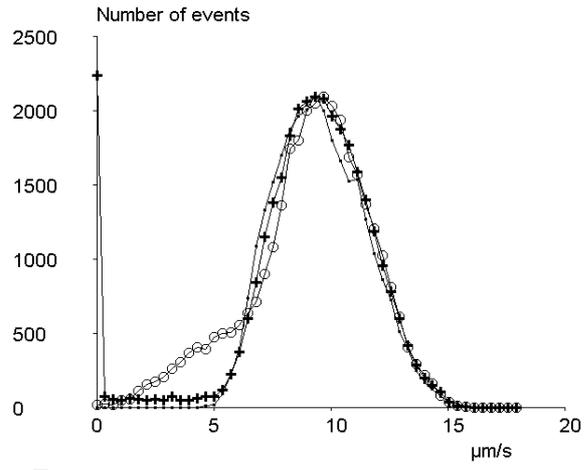

A    B

Fig. 3

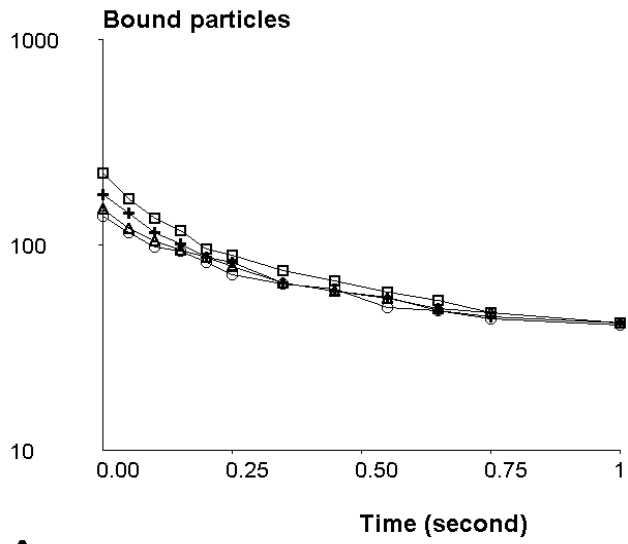
A

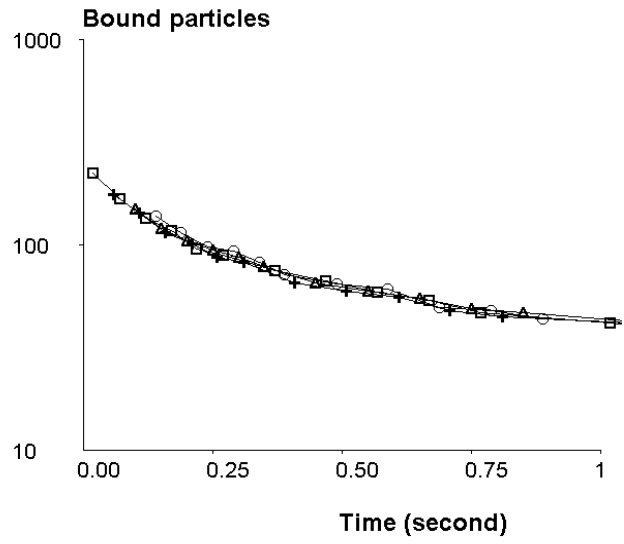
B

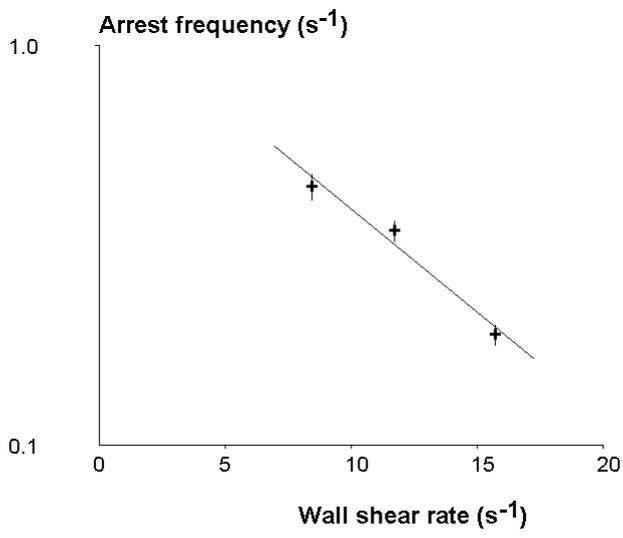
C

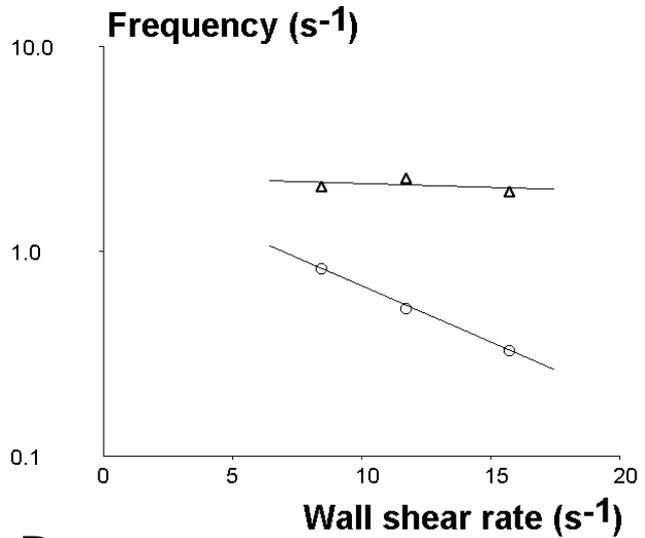
D

Fig. 4

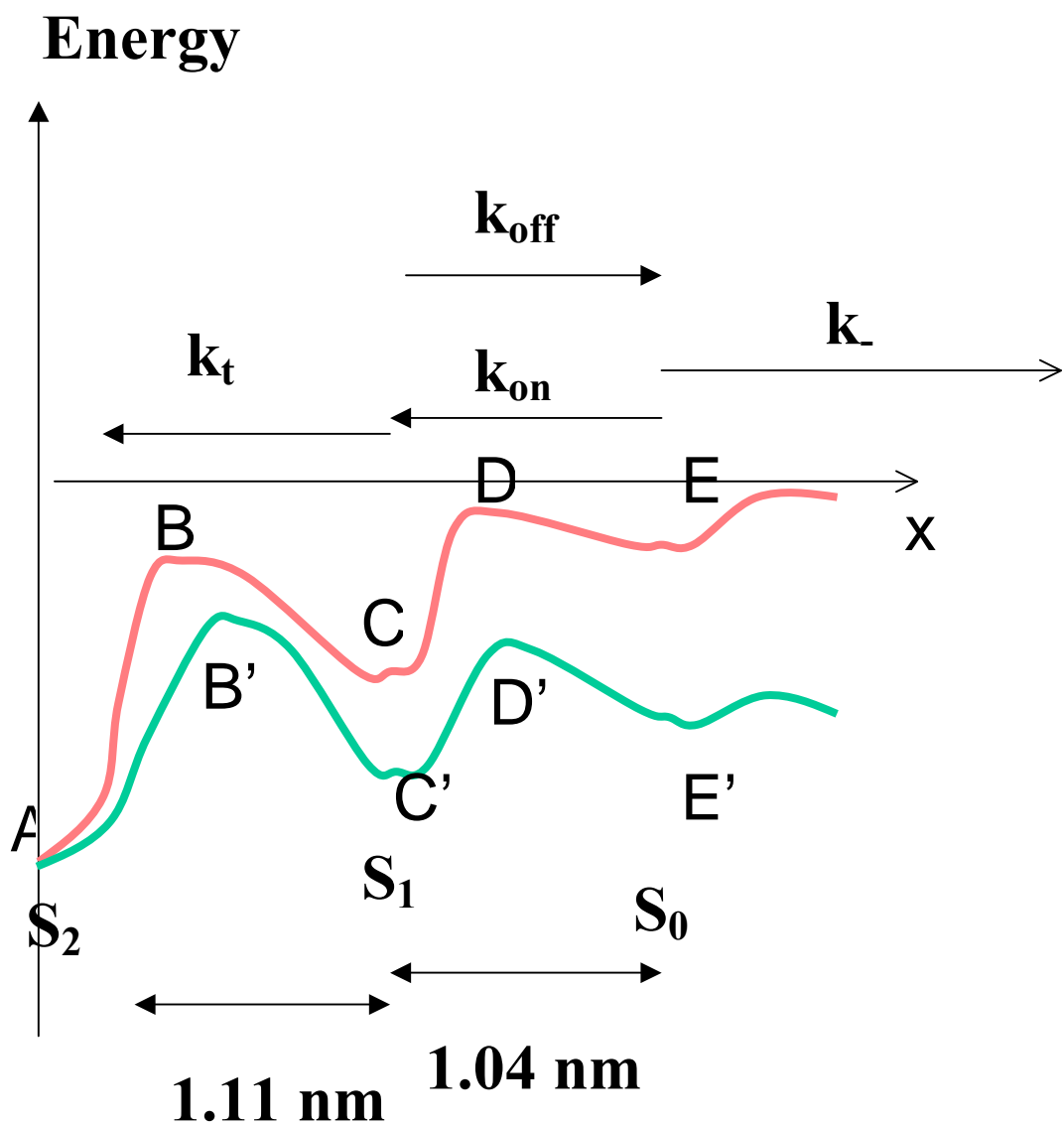